\documentclass{article}

\usepackage{arxiv}

\usepackage[utf8]{inputenc} 
\usepackage[T1]{fontenc}    
\usepackage{hyperref}       
\usepackage{url}            
\usepackage{booktabs}       
\usepackage{amsfonts}       
\usepackage{nicefrac}       
\usepackage{microtype}      
\usepackage{lipsum}		
\usepackage{graphicx}
\usepackage[numbers]{natbib}
\usepackage{doi}
\usepackage{authblk}
\usepackage{amsmath}
\usepackage{amssymb}

\title{Enforcing Equity in Neural Climate Emulators}

\author[1, 2, *, $\dagger$]{William Yik}
\author[2, 3, 4]{Sam J.~Silva}

\makeatletter
\renewcommand{\AB@affilsepx}{\quad}
\protect\Affilfont
\makeatother

\affil[1]{\footnotesize Harvey Mudd College}
\affil[2]{\footnotesize Dept. of Earth Sciences, University of Southern California}
\affil[3]{\footnotesize Dept. of Civil and Environmental Engineering, University of Southern California}

\makeatletter
\renewcommand{\AB@affilsepx}{\\ \vspace{4mm}}
\protect\Affilfont
\makeatother

\affil[4]{\footnotesize Dept. of Population and Public Health Sciences, University of Southern California}

\makeatletter
\renewcommand{\AB@affilsepx}{\quad}
\protect\Affilfont
\makeatother

\affil[*]{\footnotesize Corresponding author: yikwill@uw.edu}
\affil[$\dagger$]{\footnotesize Now at Department of Atmospheric Sciences, University of Washington}

\date{\vspace{-5mm}\today}

\hypersetup{
pdftitle={Enforcing Equity in Neural Climate Emulators},
pdfsubject={physics.ao-ph, cs.LG},
pdfauthor={William Yik, Sam J.~Silva},
pdfkeywords={AI for climate, machine learning, equity, fairness},
}

\begin{document}
\maketitle

\begin{abstract}
Neural network emulators have become an invaluable tool for a wide variety of climate and weather prediction tasks. While showing incredibly promising results, these networks do not have an inherent ability to produce equitable predictions. That is, they are not guaranteed to provide a uniform quality of prediction along any particular class or group of people. This potential for inequitable predictions motivates the need for explicit representations of fairness in these neural networks. To that end, we draw on methods for enforcing analytical physical constraints in neural networks to bias networks towards more equitable predictions. We demonstrate the promise of this methodology using the task of climate model emulation. Specifically, we propose a custom loss function which punishes emulators with unequal quality of predictions across any prespecified regions or category, here defined using human development index (HDI). This loss function weighs a standard loss metric such as mean squared error against another metric which captures inequity along the equity category (HDI), allowing us to adjust the priority of each term before training. Importantly, the loss function does not specify a particular definition of equity to bias the neural network towards, opening the door for custom fairness metrics. Our results show that neural climate emulators trained with our loss function provide more equitable predictions and that the equity metric improves with greater weighting in the loss function. We empirically demonstrate that while there is a tradeoff between accuracy and equity when prioritizing the latter during training, an appropriate selection of the equity priority hyperparameter can minimize loss of performance.
\end{abstract}

\ifx
\begin{center}
{\large \bfseries \scshape Significance}
\begin{quote}
Earth system models play a key role in our understanding of the climate and are indispensable in guiding global climate policy. Recent advances in deep learning have shown promise for enhancing or replacing these traditional models because of their low computational cost and predictive power. However, these so called neural climate emulators are almost always solely optimized to provide stable, accurate predictions at long lead times, ignoring potential inequities in the model’s performance for different regions of the globe. In this work, we propose a novel framework for training neural climate emulators which explicitly encodes fairness goals. Our results show that it is possible to bias emulators towards more equitable predictions for all, while making minimal sacrifices in model accuracy.
\end{quote}
\end{center}
\fi


\section{Introduction}
Modern Earth system models (ESMs) are key in characterizing the Earth's response to anthropogenic forcing, and their predictions have been widely used to guide global climate policy such as United Nations Paris Agreement. As such, the impact of the predictions from such models has the potential to reach every human being on Earth. By numerically solving equations which describe our understanding of the climate system, ESMs are able to predict the state of the planet under various future scenarios. The large computational cost of such ESMs \citep{collins2012quantifying,o2016scenario}, however, has motivated recent applications of new machine learning (ML) techniques for climate prediction, in particular deep learning with neural networks \citep{watson2022climatebench,kochkov2023neural,watt2023ace,el2022global,bauer2023deep,yik2023exploring}. While successive iterations of these neural climate emulators have each pushed the frontier of stable climate predictions using machine learning, the metrics used to assess these models only focus on accuracy or, more rarely, sensitivity to forcing and physical consistency/plausibility \citep{rasp2020weatherbench,rasp2023weatherbench,yu2024climsim}.

One important aspect of climate predictions which is discussed even less frequently is their \textit{fairness} or \textit{equity}. This consideration is of particular importance for emulators given their increasing application across a wide variety of climate prediction tasks including atmospheric forecasting \citep{watt2023ace,pathak2022fourcastnet,kochkov2023neural}, subgrid-scale parameterization \citep{wang2022non,rasp2018deep,beucler2024climate,zanna2021deep}, precipitation nowcasting \citep{shi2017deep,espeholt2022deep,li2022using}, equation and knowledge discovery \citep{zanna2020data,grundner2024data,yik2023southern}, data assimilation \citep{arcucci2021deep,wang2022deep}, downscaling \citep{wang2022down,harder2023hard,geiss2022downscaling}, bias correction \citep{han2021deep,hess2023deep,kim2021deep}, and more \citep{lai2024machine}. Such principles also align themselves with the explicit fairness goals of many climate policies as well as recent literature on policy development \citep{rogna2022optimal,giang2024equity}. Despite this, there has been little work to date discussing equity in neural climate emulator predictions. Nevertheless, the inherent process of training a neural network provides a unique opportunity to address this issue. Since researchers already define advanced loss functions to bias their emulators towards specific goals, such as long-term stability and forecast sharpness \citep{pathak2022fourcastnet,lam2023learning}, the loss function itself could open a path towards more equitable predictions if notions of fairness could be integrated within it.

In this work, we draw on methods for enforcing physical constraints (e.g., conservation of mass and energy) \citep{beucler2021enforcing} in neural networks via the loss function to bias neural climate emulators of global temperature towards more equitable predictions. Specifically, we use a two-part loss function which weighs a standard error metric such as MSE against a measure of the model's fairness on a sliding scale. In doing so we can adjust the model's preference for predictions with equal error throughout different prespecified regions of the globe, defined here using Human Development Index \citep{anand1994human}. Importantly, given the many existing quantitative measures of equity, our method does not specify a specific fairness metric and allows for any such metric to be weighed against error in the loss function. This is in contrast to previous methods which focus specifically on spatial biases in neural network predictions \citep{he2023physics} or non-spatial tabular data \citep{perez2017fair}. To demonstrate our method's capability, we train a neural climate emulator to predict surface air temperature and the diurnal temperature range through the year 2100 using our custom loss function. We show that emulators trained using the loss function behave as desired, providing more equitable predictions and improving the equity metric as it is given greater weighting in the loss function. Additionally, we demonstrate that the sliding scale nature of the loss function can give rise to a tradeoff between accuracy and equity when the latter is weighted heavily during training. We demonstrate that a sufficiently small choice of equity weighting could vastly improve the fairness metric while achieving only a small loss in prediction accuracy.

\section{Equitable Loss Functions}\label{sec: equitable loss functions}
The ultimate goal of this work is to bias neural climate emulators towards more equitable predictions by integrating notions of fairness into their objective functions. This is accomplished in a manner similar to that of previous methods for enforcing analytical physical constraints in neural networks \citep{beucler2021enforcing,harder2022physics,harder2023hard}. The key idea is to balance a traditional measure of error, in our case mean squared error, against a penalty term which captures some representation of fairness in the neural networks. Specifically, the loss function is formulated as
\begin{align*}
    \mathcal{L} = \alpha \mathcal{P} + (1-\alpha)\text{MSE},
\end{align*}
where $\mathcal{P}$ is the equity penalty, MSE is global mean squared error, and $\alpha$ is the equity weighting coefficient. Under this construction, $\mathcal{P}$ may be based on any quantitative fairness metric. Given the lack of a singular definition for fairness, this flexibility is particularly important. We illustrate the effect of our loss function by defining fairness based on prediction accuracy in various predefined regions across the globe. Specifically, we divide the land of Earth into $n$ equally sized regions based on Human Development Index, a composite measure of a country's development \citep{anand1994human,kummu2018gridded}. Then for the term $\mathcal{P}$, we punish the \textit{coefficient of variation}, or relative standard deviation, between the errors across those $n$ regions, encouraging the emulator to have equal predictive error in each region. While this is the only equity penalty presented in this work, we found that other definitions such as simple standard deviation and performance deviation from the most well predicted region lead to similar conclusions, with modestly worse overall performance.

\begin{figure}[t]
    \centering
    \includegraphics[width=\columnwidth]{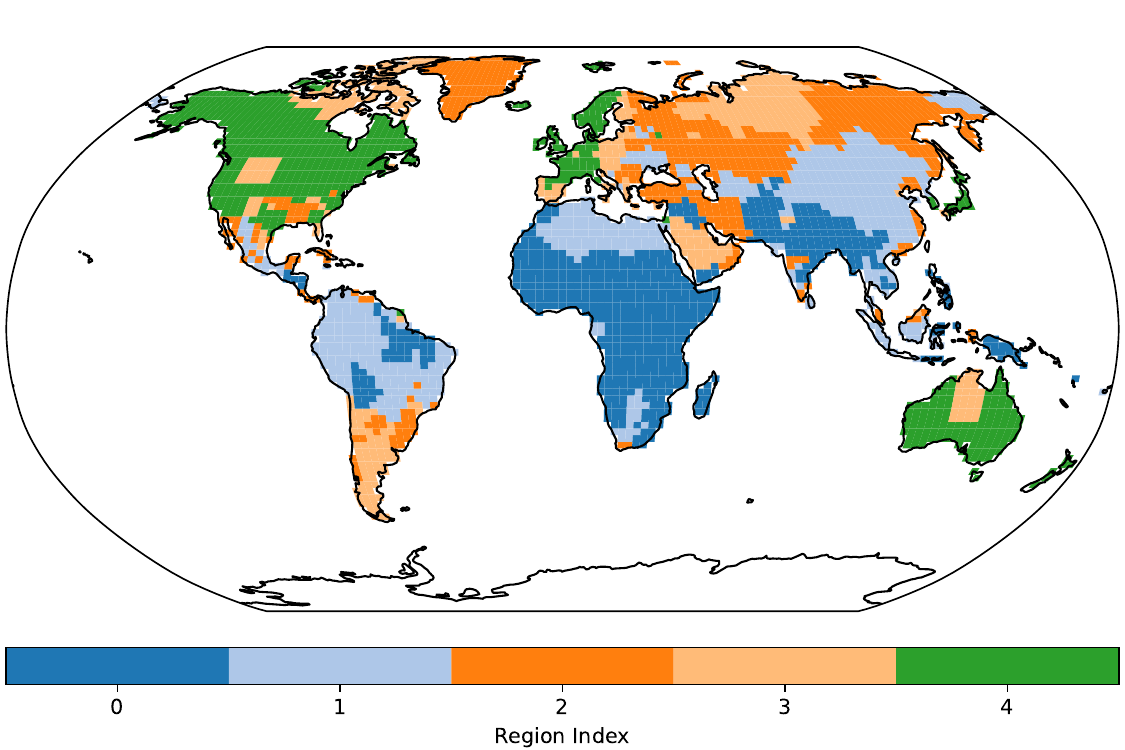}
    \caption{The 5 HDI regions over which equity is defined in the penalty $\mathcal{P}$.}
    \label{fig: hdi regions}
\end{figure}

The particular penalty explored in this work seeks to equalize the loss function for different groups, which is akin to parity-based fairness metrics in statistical applications \citep{agarwal2019fair,caton2020fairness}. More precisely, if $\text{MSE}_i$ is the MSE of the emulator's prediction in region $i$, then the emulator's MSE over all land is
\begin{align*}
    \text{MSE}_l = \frac{1}{n}\sum_{i=0}^{n-1}\text{MSE}_i.
\end{align*}
Furthermore, the standard deviation of the MSE's for each region is
\begin{align*}
    \sigma_{R} = \sqrt{\frac{\sum_{i=0}^{n-1}(\text{MSE}_i-\text{MSE}_l)^2}{n}}.
\end{align*}
Finally, our equity penalty for the climate emulator is
\begin{align*}
     \mathcal{P} = \frac{\sigma_R}{\text{MSE}_l}.
\end{align*}
We split the land grid cells into quantiles each containing 20\% of the cells (i.e., $n=5$) based on the prespecified equity measure of interest, HDI. The spatial distribution of these 5 HDI regions are shown in Fig. \ref{fig: hdi regions}. 

\section{Results}
To demonstrate our equitable loss function methodology, we train an ensemble of neural networks to predict one of either surface air temperature (TAS) or diurnal temperature range (DTR) for the years 2080-2100 under an intermediate climate forcing scenario \cite{o2016scenario} using the ClimateBench dataset \cite{watson2022climatebench}. We emphasize again, however, that our method is agnostic to the training dataset and chosen definition of quantitative fairness. These neural climate emulators take as input annual means of carbon dioxide (CO$_2$), methane (CH$_4$), sulfur dioxide (SO$_2$), and black carbon (BC) on a 96 latitude $\times$ 144 longitude global grid and predict TAS and DTR at the same spatiotemporal resolution. (See \textit{Materials and Methods} for further details.)

\begin{figure*}[t]
    \begin{minipage}[t]{0.5\linewidth}
        \includegraphics[width=\linewidth]{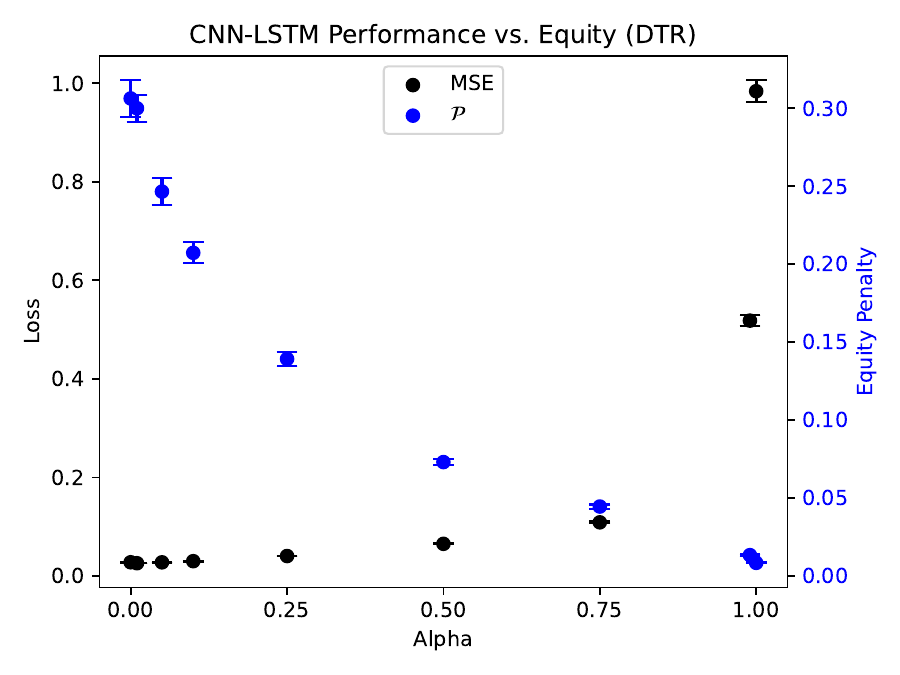}
        \caption{MSE (black) and $\mathcal{P}$ (blue) for each neural network ensemble trained to predict DTR with varying $\alpha$. Error bars represent standard error of the ensemble mean.}
        \label{fig: cvpenalty performance equity dtr}
    \end{minipage}%
    \hfill
    \begin{minipage}[t]{0.5\linewidth}
        \includegraphics[width=\linewidth]{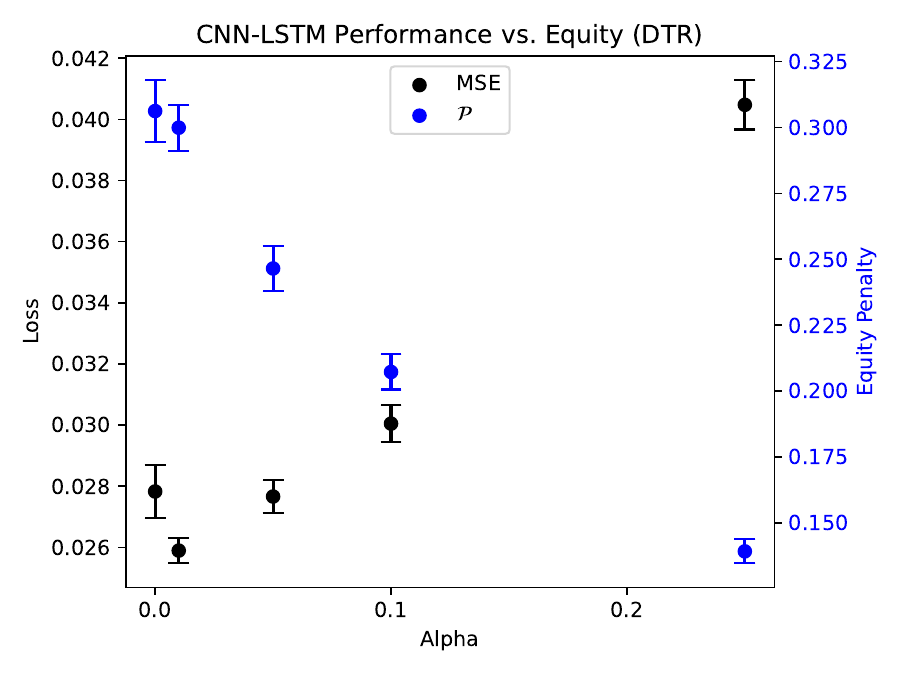}
        \caption{As in Fig. \ref{fig: cvpenalty performance equity dtr}, but only for $\alpha \leq 0.25$.}
        \label{fig: cvpenalty performance equity dtr zoomed}
    \end{minipage}
    \begin{minipage}[t]{0.5\linewidth}
        \includegraphics[width=\linewidth]{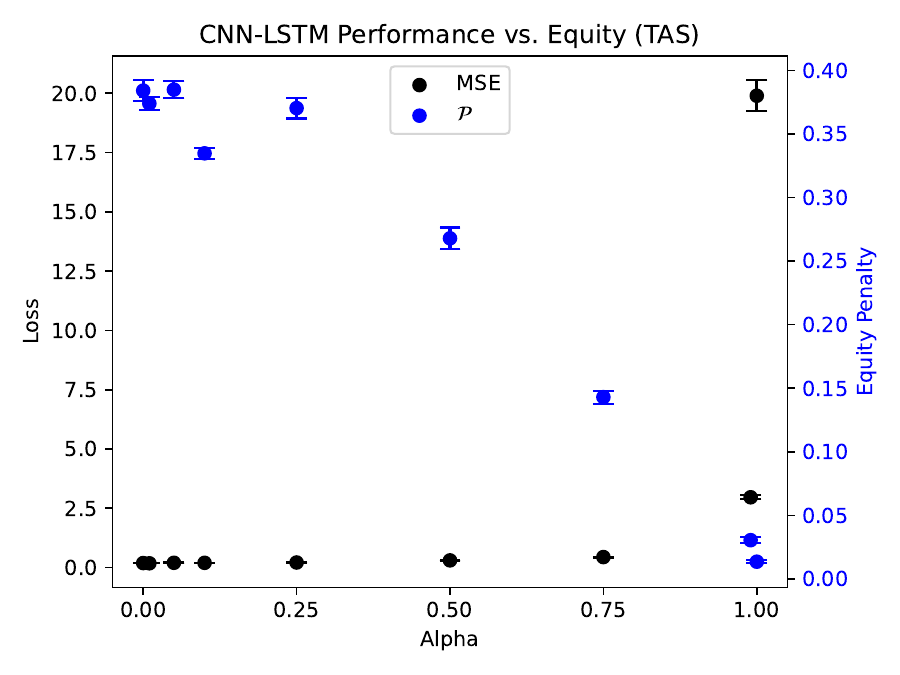}
        \caption{MSE (black) and $\mathcal{P}$ (blue) for each neural network ensemble trained to predict TAS with varying $\alpha$. Error bars represent standard error of the ensemble mean.}
        \label{fig: cvpenalty performance equity tas}
    \end{minipage}%
    \hfill
    \begin{minipage}[t]{0.5\linewidth}
        \includegraphics[width=\linewidth]{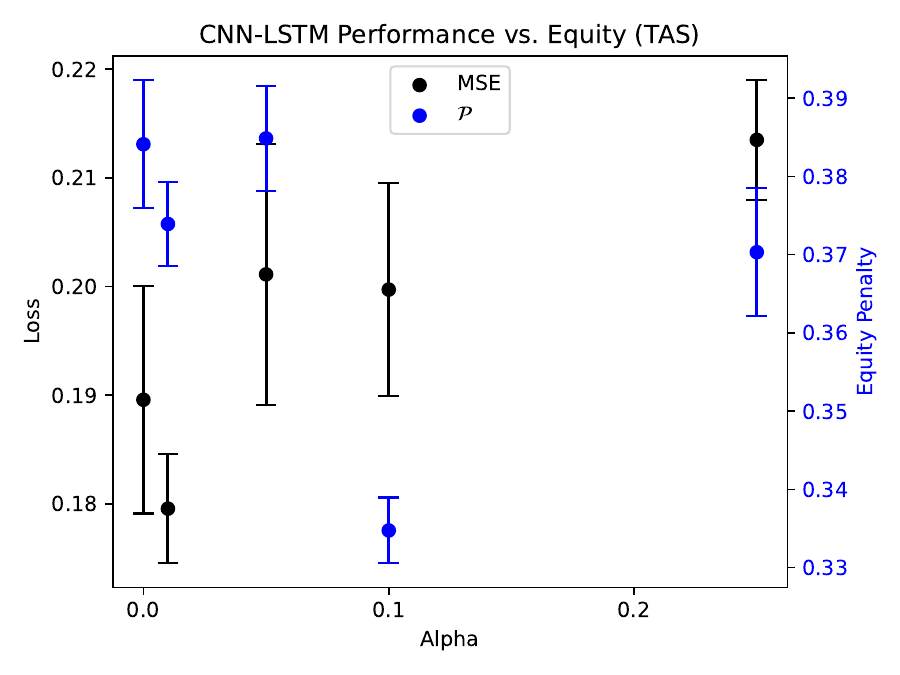}
        \caption{As in Fig. \ref{fig: cvpenalty performance equity tas}, but only for $\alpha \leq 0.25$.}
        \label{fig: cvpenalty performance equity tas zoomed}
    \end{minipage}
\end{figure*}

We generate  ensembles of 15 models for each of  $\alpha=0,0.01,0.05,0.1,0.25,0.5,0.75,$ and $1$. Then we compute the MSE of each trained ensemble member's prediction, as well as the equity penalty. The results are shown in Fig. \ref{fig: cvpenalty performance equity dtr}. Immediately obvious is the near monotonic increasing trend for MSE with increasing $\alpha$ and a corresponding near monotonic decreasing trend for the equity penalty $\mathcal{P}$. Given the construction of the loss function with equity weighting coefficient $\alpha$, this tradeoff between accuracy and fairness is expected. Interestingly, for low values of $\alpha$, there appear to be large relative gains in the equity penalty in exchange for very little relative loss in predictive power. This is highlighted in Fig. \ref{fig: cvpenalty performance equity dtr zoomed} which shows the same plot as Fig. \ref{fig: cvpenalty performance equity dtr} but only for $\alpha \leq 0.25$. For example, compared to $\alpha=0$, the ensemble trained with $\alpha=0.1$ sees a 32\% improvement in the equity penalty for only an 8\% loss in accuracy.

Figs. \ref{fig: cvpenalty performance equity tas} and \ref{fig: cvpenalty performance equity tas zoomed} show a similar tradeoff between performance and equity for the TAS prediction task. It seems, however, that the $\alpha$ value for which large gains in the equity penalty come at low cost in performance is between $\alpha=0.25$ and $\alpha=0.5$, as opposed to around $\alpha=0.1$ for DTR. We do not necessarily expect that this ``sweet spot" value for $\alpha$ should be the same for both prediction tasks, as the errors for each will have different scales which can affect both the MSE and $\mathcal{P}$ values in the loss function.

\begin{figure}[t!]
    \centering
    \includegraphics[width=\columnwidth]{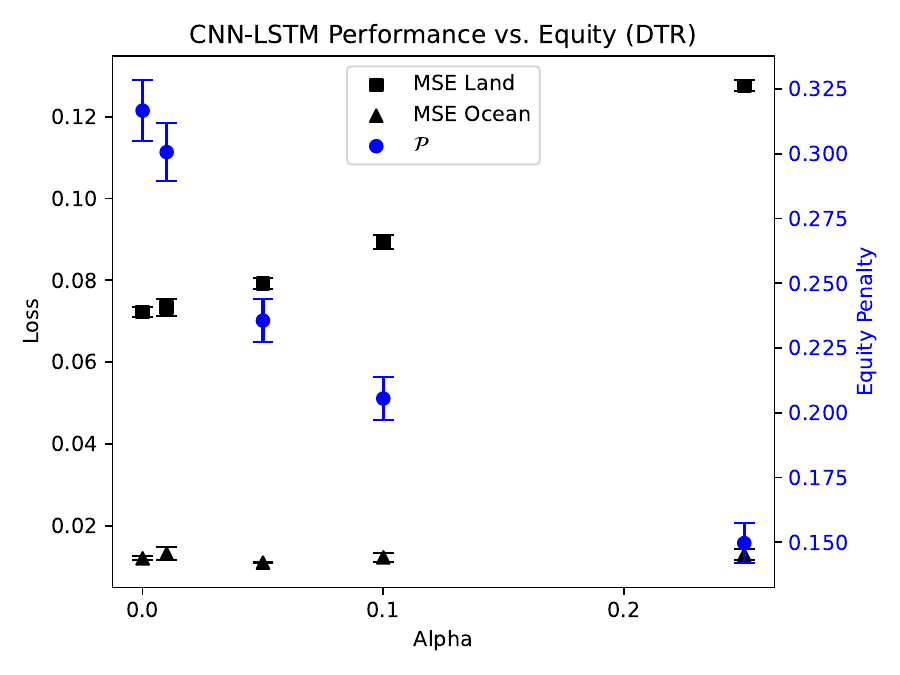}
    \caption{MSE (black) and $\mathcal{P}$ (blue) for each neural network ensemble trained to predict DTR with varying $\alpha$. The square and triangular points represent MSE over land and ocean, respectfully. Error bars represent standard error of the ensemble mean.}
    \label{fig: cvpenalty land sea performance equity dtr}
\end{figure}

For both DTR and TAS, certain ensembles trained with low $\alpha$ values exhibit interesting behavior where both the accuracy and equity penalty of the ensemble are modestly improved compared to the $\alpha=0$ ensemble, though the $\alpha$ value where this occurs is not consistent between DTR and TAS. This is especially surprising since the MSE component of the loss function is defined globally while the equity penalty is only defined over land, meaning that the emulator is not simply achieving more equitable predictions by significantly raising error over the ocean. This is illustrated in Fig. \ref{fig: cvpenalty land sea performance equity dtr} which shows the emulator's MSE over both land and sea, as well as the equity penalty, for $\alpha \leq 0.25$. From this, it is seen that the emulator's performance over the ocean does not degrade with increasing $\alpha$ and in fact improves for certain values.

Overall, Fig. \ref{fig: cvpenalty performance equity dtr}-\ref{fig: cvpenalty performance equity tas zoomed} highlight that it is indeed possible to bias neural climate emulators towards more equitable predictions with a loss function that explicitly encodes a fairness metric. While this comes at the cost of accuracy, an appropriate choice of a small equity weighting coefficient $\alpha$ (here, $\alpha \lessapprox 0.25$) can mitigate losses in performance or even improve global predictions, narrowing the $\alpha$ hyperparameter search space for those wishing to make use of equitable loss functions. This finding for low $\alpha$ values aligns with those of prior work \cite{beucler2021enforcing} for enforcing analytical physical constraints in neural networks.

\begin{figure}[t!]
    \begin{minipage}[t]{\linewidth}
        \includegraphics[width=\linewidth]{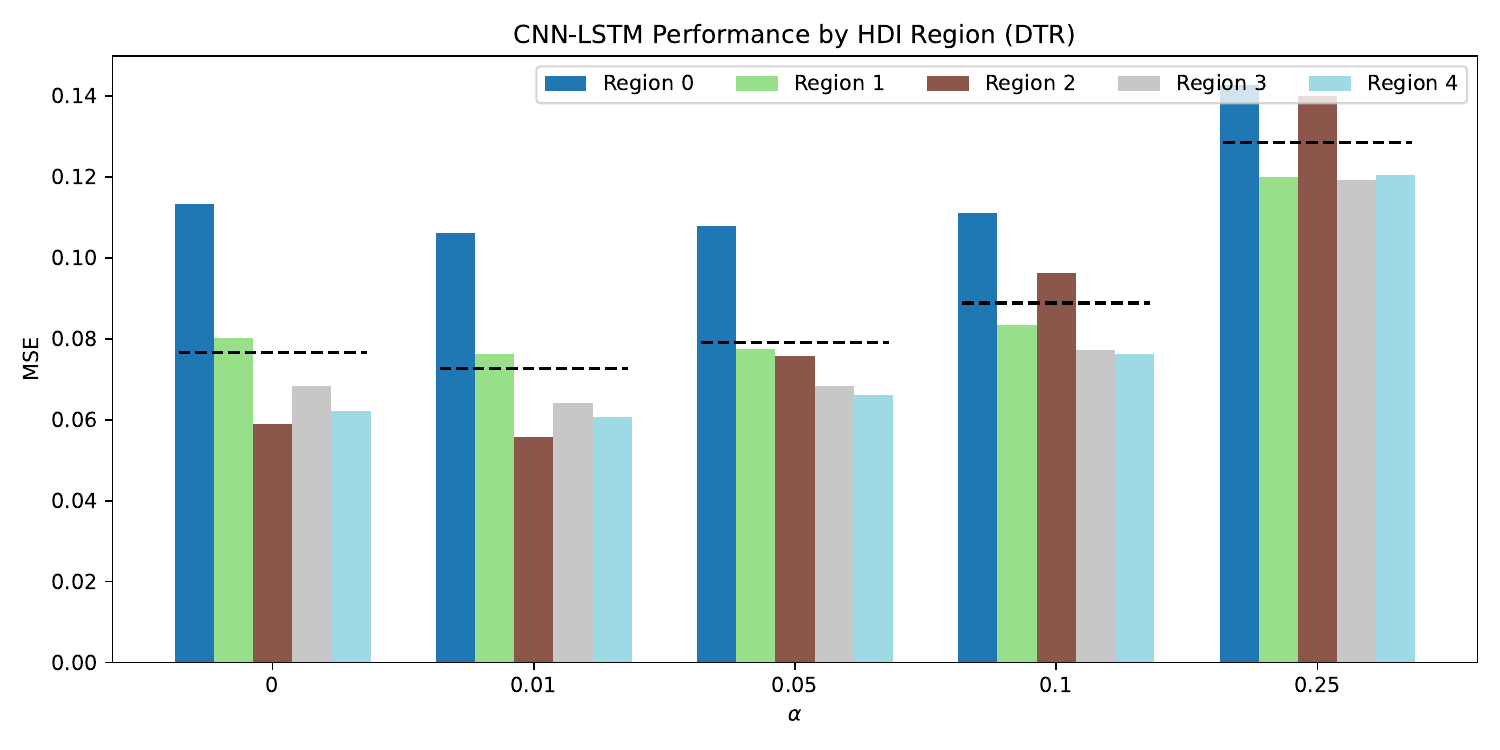}
        \caption{DTR MSE in each HDI region for several values of $\alpha$. The black dashed line shows the mean of the colored bars, or simply the MSE over land.}
        \label{fig: cvpenalty error per region dtr}
    \end{minipage}
    \hfill
    \begin{minipage}[t]{\linewidth}
        \includegraphics[width=\linewidth]{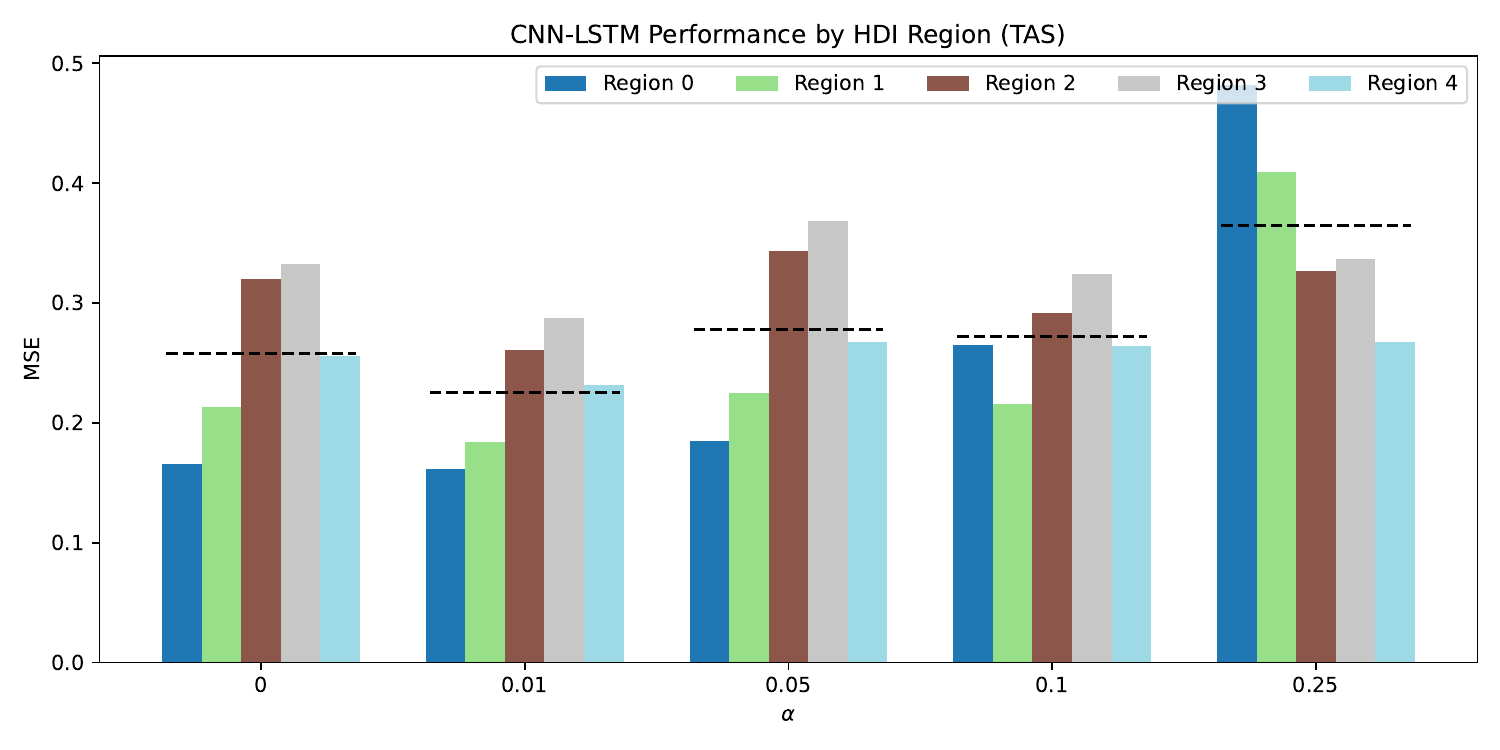}
        \caption{As in Fig. \ref{fig: cvpenalty error per region dtr}, but TAS.}
        \label{fig: cvpenalty error per region tas}
    \end{minipage}
\end{figure}

\begin{figure}[t!]
    \centering
    \includegraphics[width=\columnwidth]{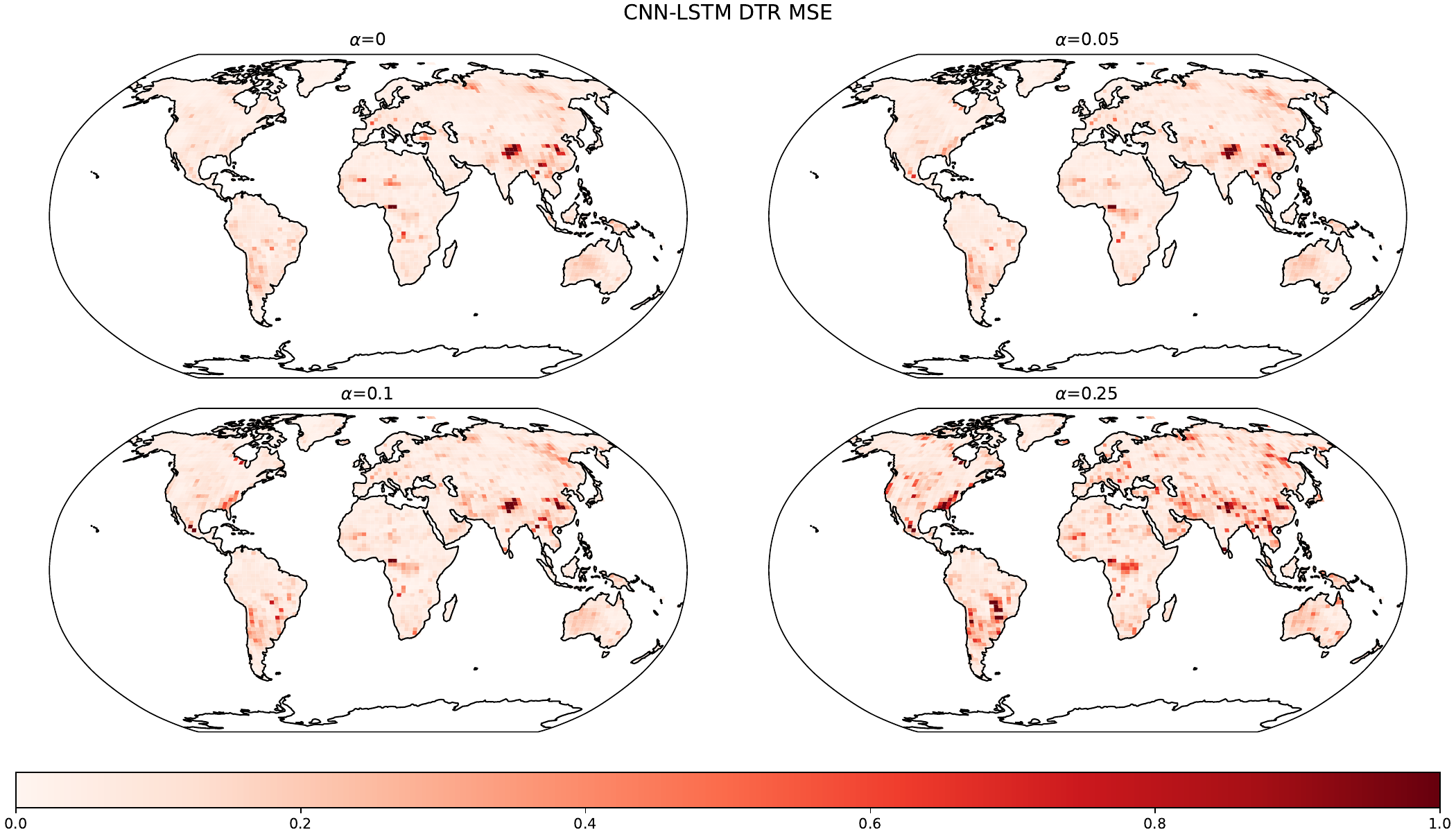}
    \caption{MSE of the neural network's DTR predictions averaged over the test data for $\alpha=0$, $0.05$, $0.1$, and $0.25$.}
    \label{fig: global error dtr}
\end{figure}

To further understand the performance of neural climate emulators trained with the equitable loss function, we break down the predictive error at several small $\alpha$ values by the HDI regions (see Figure 1). This is illustrated in Fig. \ref{fig: cvpenalty error per region dtr}, which shows the emulator's predictive error in each HDI region for five increasing values of $\alpha$ (0, 0.01, 0.05, 0.1, and 0.25) for DTR, and Fig. \ref{fig: cvpenalty error per region tas}, which shows the same results but for the set of emulators trained to predict TAS. These results highlight how error is redistributed between the HDI regions in order to lower the equity penalty as $\alpha$ increases. Specifically, the ensemble achieves more equitable predictions at higher values of $\alpha$ by lowering the relative deviation of the error in each region from the mean (black dashed line). This is accomplished by simultaneously lowering the error in poorly predicted regions and raising the error in well predicted regions. For both predicted variables, the $\alpha=0.1$ neural network ensemble achieves similar MSE performnce as the $\alpha=0$ case while also successfully reducing the spread in error between the HDI regions.

By redistributing error throughout the HDI regions as shown in Figs. \ref{fig: cvpenalty error per region dtr} and \ref{fig: cvpenalty error per region tas}, neural climate emulators trained with the equitable loss function make the spatial distribution of predictive error more uniform. That is, they make predictions of equal quality throughout the globe. This is illustrated in Fig. \ref{fig: global error dtr} which shows the DTR MSE averaged over the test period (2080-2100) for four increasing values of $\alpha$ (0, 0.05, 0.1, and 0.25). Most obviously for the $\alpha=0.25$ ensemble compared to $\alpha=0$, the error in the poorly predicted regions over the Tibetan plateau and southeast Asia decreases while error in other regions of the globe increases, effectively lowering the equity penalty. For the regions where predictions become worse at higher $\alpha$ values, the ensemble appears to distribute error randomly within them. This trend towards a random uniform distribution of error across the globe becomes more and more obvious with increasing $\alpha$.

\section{Discussion and Conclusion}\label{sec: discussion and conclusion}
As weather and climate predictions from both pure ML and hybrid physics-ML models continue to improve according to global accuracy metrics, it remains equally important to ensure that their predictions are accurate for everyone, regardless of their location on the globe. By utilizing a novel loss function which balances a traditional error metric with a penalty for inequitable predictions, our method shows that neural climate emulators can be biased towards more fair predictions, regardless of the chosen definition of quantitative equity. Moreover, with proper weighting of the equity penalty term in the loss function, this comes at little cost to global accuracy. Overall, our results highlight that neural climate emulators can achieve fair predictions when trained towards an appropriate goal.

While we focus on one particular definition of equity in this work, another may easily be substituted in this proposed loss function penalty framework. There are interesting implications for the loss landscape when using different definitions of numerical equity. Just as mean squared error is preferred over mean absolute error when optimizing neural networks for certain problems \citep{hodson2022root,chai2014root}, some equity penalties will be more favorable from an optimization standpoint than others. Similarly, our choice to define equity regions based on HDI is not the only possible option. A more spatially coherent choice (e.g., using the continents as regions) is easily achievable using our method, and there are interesting open questions regarding the effects this may have on training and performance. Furthermore, many recent AI weather models tend to be optimized towards multi-objective loss functions, often to achieve accuracy on both short and long lead times. In a similar manner, future work may investigate the effects of training a neural climate emulator towards both equity and another goals such as obeying a physical constraint or maintaining sharpness in predictions at long lead times.

\section{Materials and Methods}
\subsection{Data}
We train neural networks to emulate the outputs of the Norwegian Earth System Model version 2 (NorESM2) \citep{seland2020overview} using the ClimateBench dataset \citep{watson2022climatebench}. For several shared socioeconomic pathways (SSPs) covering a wide range of future emissions scenarios \citep{o2016scenario}, the ClimateBench dataset provides four key inputs to NorESM2, namely carbon dioxide (CO$_2$), methane (CH$_4$), sulfur dioxide (SO$_2$), and black carbon (BC). These inputs are given as annual means on a 96 latitude $\times$ 144 longitude global grid. We use these inputs to predict the surface air temperature (TAS) and diurnal temperature range (DTR) outputs of NorESM2 at the same 96 latitude $\times$ 144 longitude global grid spatiotemporal resolution.

\subsection{Neural Network Model and Training}
In this work, we use the best performing model analyzed in \cite{watson2022climatebench}, a CNN-LSTM. This type of neural network combines a convolutional neural network (CNN) with a long short-term memory (LSTM) network, which respectively capture spatial and temporal relations in the data making them well suited for climate prediction.

For each of the temperature variables we wish to predict (TAS and DTR) and $\alpha$ value explored ($0,0.01,0.05,0.1,0.25,0.5,0.75,$ and $1$), we train an ensemble of 15 CNN-LSTMs on data provided by ClimateBench for a total of $2 \times 8 \times 15 = 240$ neural networks. The training data includes NorESM2's historical experiment as well as its output for the SSP126, SSP370, and SSP585 experiments of the Scenario Model Intercomparison Project (ScenarioMIP) \cite{o2016scenario}. We also include the hist-GHG and hist-aer experiments from the Detection and Attribution Model Intercomparison Project (DAMIP) \citep{gillett2016detection} in our training data. Each of the 3 SSP and 3 historical experiments contain data from 2015-2100 and 1850-2014 for a total of 753 training data points. From this training dataset, we reserve the first two years of data from every decade in each experiment as validation data. This is done because climate data is highly autocorrelated in time, so it best practice to form the validation data from continuous subsets of dataset rather than at random. Finally, to evaluate the CNN-LSTM's performance, we use the SSP245 scenario as a holdout test dataset because it represents intermediate future anthropogenic forcing.

\subsection{Data, Materials, and Software Availability} The complete codebase for processing the ClimateBench data, defining the equitable loss function, training the neural networks, and analyzing the results will be made available soon.

\section*{Acknowledgements}
The authors thank Joseph Hardin for helpful discussions and feedback.

\bibliographystyle{unsrtnat}
\bibliography{references}  






\end{document}